\documentclass[fleqn,10pt]{wlscirep}
\usepackage[utf8]{inputenc}
\usepackage[T1]{fontenc}
\usepackage{amsmath}
\usepackage{enumitem}

\title{Probability density function for random photon steps in a binary (isotropic-Poisson) statistical mixture}

\author[1,*]{Tiziano Binzoni}
\author[2]{Alain Mazzolo}

\affil[1]{Department of Radiology and Medical Informatics, University Hospital, Geneva, 1211, Switzerland}
\affil[2]{Universit\'e Paris-Saclay, CEA, Service d'\'Etudes des R\'eacteurs et de Math\'ematiques Appliqu\'ees, 91191, Gif-sur-Yvette, France}

\affil[*]{tiziano.binzoni@unige.ch}

\begin{abstract}
Monte Carlo (MC) simulations allowing to describe photons propagation in statistical mixtures represent an interest that 
goes way beyond the domain of optics, and can cover, e.g., nuclear reactor physics,  image analysis  or life science just to name a few.
MC simulations are considered a ``gold standard'' because  they give  exact solutions (in the statistical sense), 
 however,  in the case of statistical mixtures they are enormously time consuming and their implementation is often extremely complex.
For this reason, the aim of the present contribution is to propose a new approach that should allow us in the future to simplify the MC approach. 
This is done through an explanatory example, i.e.; by deriving the `exact' analytical expression for the probability density function of photons' random 
steps (single step function, SSF) propagating in a medium represented as a binary (isotropic-Poisson) statistical mixture. 
The use of the  SSF  reduces the problem  to an `equivalent'  homogeneous medium behaving exactly as the original binary statistical mixture.
This will reduce hundreds time-consuming MC simulations to only one equivalent simple MC simulation.
To the best of our knowledge the analytically `exact' SSF for a binary (isotropic-Poisson) statistical mixture has never been derived before.
\end{abstract}

\begin{document}

\flushbottom
\maketitle
%
%
\thispagestyle{empty}

\section*{Introduction}

Two probability density functions (pdf) are at the heart of Monte Carlo simulations describing photon propagation in biomedical optics, optics or, in general, particle transport in diffusive media: the ``phase function'' (PF) and the pdf allowing to describe the probability for a photon to reach a distance 
$s\in [s-\frac{ds}{2},s+\frac{ds}{2}]$ 
in the medium, without interactions.
For simplicity, we will call the latter pdf ``single step function'' (SSF). 
The knowledge of the SSF is fundamental because 
it is mandatory for the implementation of a  MC simulation, and because 
from the SSF we can extract the main characteristics of photon propagation through the 
medium \cite{ref:BinzoniMartelli2022}.

The specific shape of the SSF is determined, in a very complex way, by the physical characteristics of the investigated media.
For this reason, in the majority of the cases, the SSF cannot be derived theoretically, but is estimated from experimental data. 
Among the  different SSFs appearing in the literature, one in particular seems to be more recurrent and applicable in many situations; i.e.,
the one derived from the so called Beer--Lambert--Bouguer law, with pdf
\begin{equation}
p_{LB}(s;\mu_t)=\mu_te^{-\mu_t s},
\label{eq:pLB}
\end{equation}
where $\mu_t$ is the extinction coefficient of the medium.
Historically, the SSF $p_{LB}(.)$ has been determined experimentally and, nowadays, its applicability covers
a panoply of different domains \cite{ref:atkins2018,ref:MeyerhoferEtAl2020,ref:OshinaSpigulis2021}.

\begin{figure}[ht]
\centering
\includegraphics[width=0.7\linewidth]{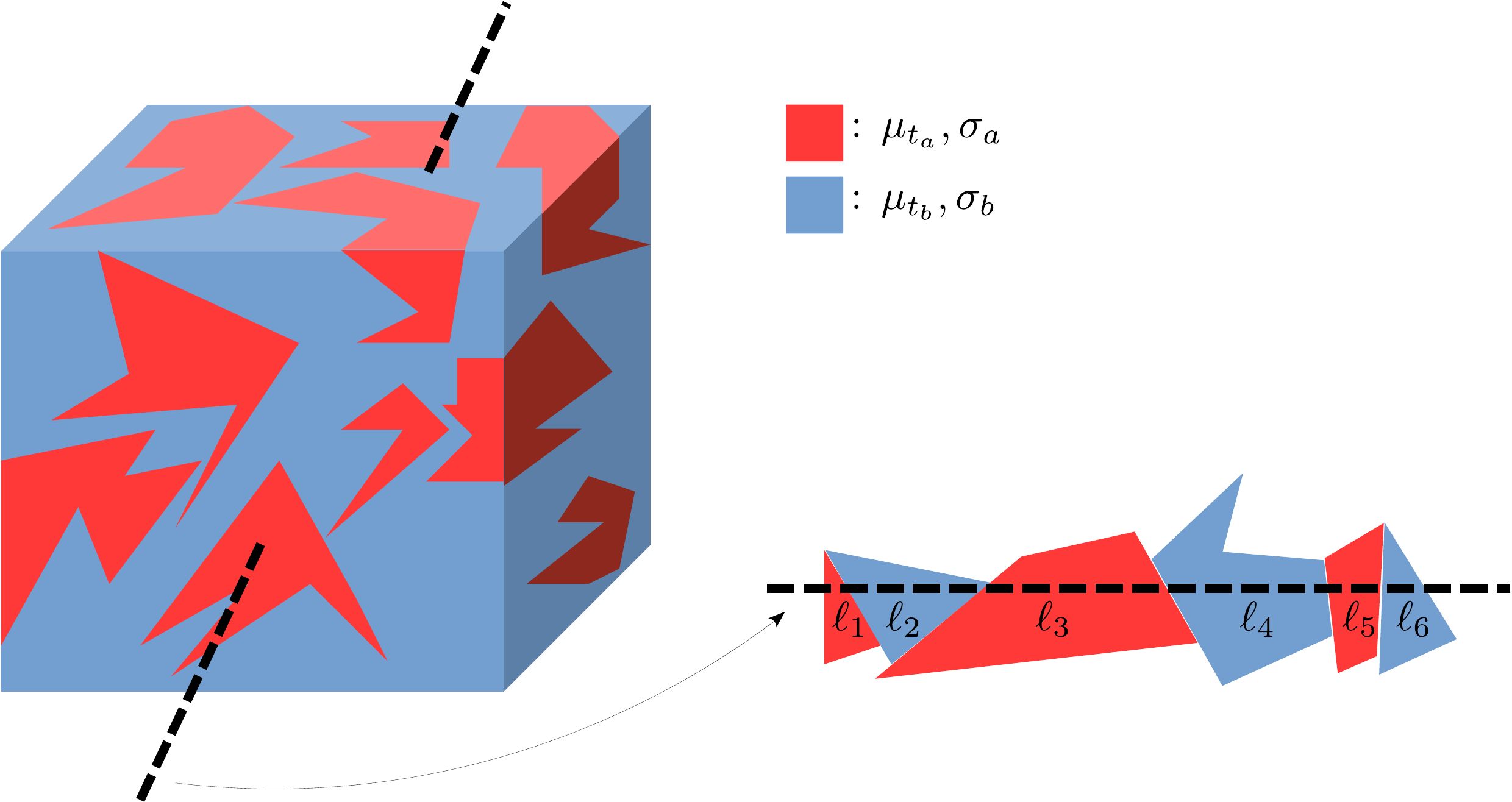}
\caption{Schematic of a 3D medium (the cubic shape is just for simplicity, and may also be infinite) composed by random tessels  of two kinds of materials  (red and sky blue) satisfying both Eq~(\ref{eq:pLB}), with $\mu_{t_a}$ and $\mu_{t_b}$.
The random oriented dashed line crosses different areas of random lengths $\ell_1, \ell_1,\dots,\ell_6$ 
(along the line), satisfying Eq.~(\ref{eq:pdfTesselLength}), with parameters $\sigma_a$ and $\sigma_b$.
If Eq.~(\ref{eq:pdfTesselLength}) with  parameters $\sigma_a$ and $\sigma_b$ always holds
for any dashed random oriented line, the medium is called isotropic.}
\label{fig:3Dgeometry-crop}
\end{figure}

Due to the large number of physical systems that can be described by $p_{LB}(.)$,  
some caution may be natural when describing systems with an SSF that appears to deviate from the Eq.~(\ref{eq:pLB}).
In fact, the question may arise whether SSFs different from Eq.~(\ref{eq:pLB})
are not simply the result generated by compound media of immiscible materials; 
where each bunch of material --- considering the subject to come in the manuscript, we will call them ``tessels'' ---
always satisfies  the classical 
Beer--Lambert--Bouguer law [Eq.~(\ref{eq:pLB})], with their own $\mu_t$
(for an intuitive example see the schematic in Fig.~\ref{fig:3Dgeometry-crop}).
With this question in mind, many studies have already been proposed in the literature; going from numerical MC simulations to theoretical approaches
\cite{ref:Pomraning1991,ref:Larmier2017,ref:Larmier2018b,ref:Larmier2018a}.  Interestingly, this approach also applies to image rendering \cite{ref:dEon2022}.

Due to their ability to describe actual physical systems, a particular attention has been given to ``random'' media obeying  mixing statistics.
In intuitive words, random media similar to the one appearing in Fig.~\ref{fig:3Dgeometry-crop}
were generated with the desired statistical law. 
Then, the exact or approximated physical quantities of interest (e.g., in the present context the  SSF)
were obtained (theoretically or numerically) on many random realizations of the medium and, finally, the ``average value'' of each quantity was derived. 
Unfortunately, the common characteristic of these studies is that they never propose    an 
{\it exact} analytical solution for the SSF.
This is why, the aim of the present contribution is to derive an exact analytical solution for 
a well celebrated model: the binary (isotropic-Poisson) statistical mixture.

But, why the knowledge of an exact analytical expression for the SSF may represent 
any advantage?
One of the reasons is that repeated MC simulations of random media,  sometimes composed by thousands 
of tessels of complex shapes, may be extremely computational expensive; in particular if simulations must be repeated a    large number of times.
The knowledge of an analytical expression for SSF, describing our random medium with mixing statistics,  allows one, in principle, to perform {\it only one simulation} on a equivalent {\it homogeneous medium}, and to obtain the same   desired  quantities
(e.g., transmitted or reflected photon fluxes) as for the original problem.

\section*{Theory}

\subsection*{Isotropic Poisson tesselation with a binary mixture}

\begin{figure}[ht]
\centering
\includegraphics[width=0.3\linewidth]{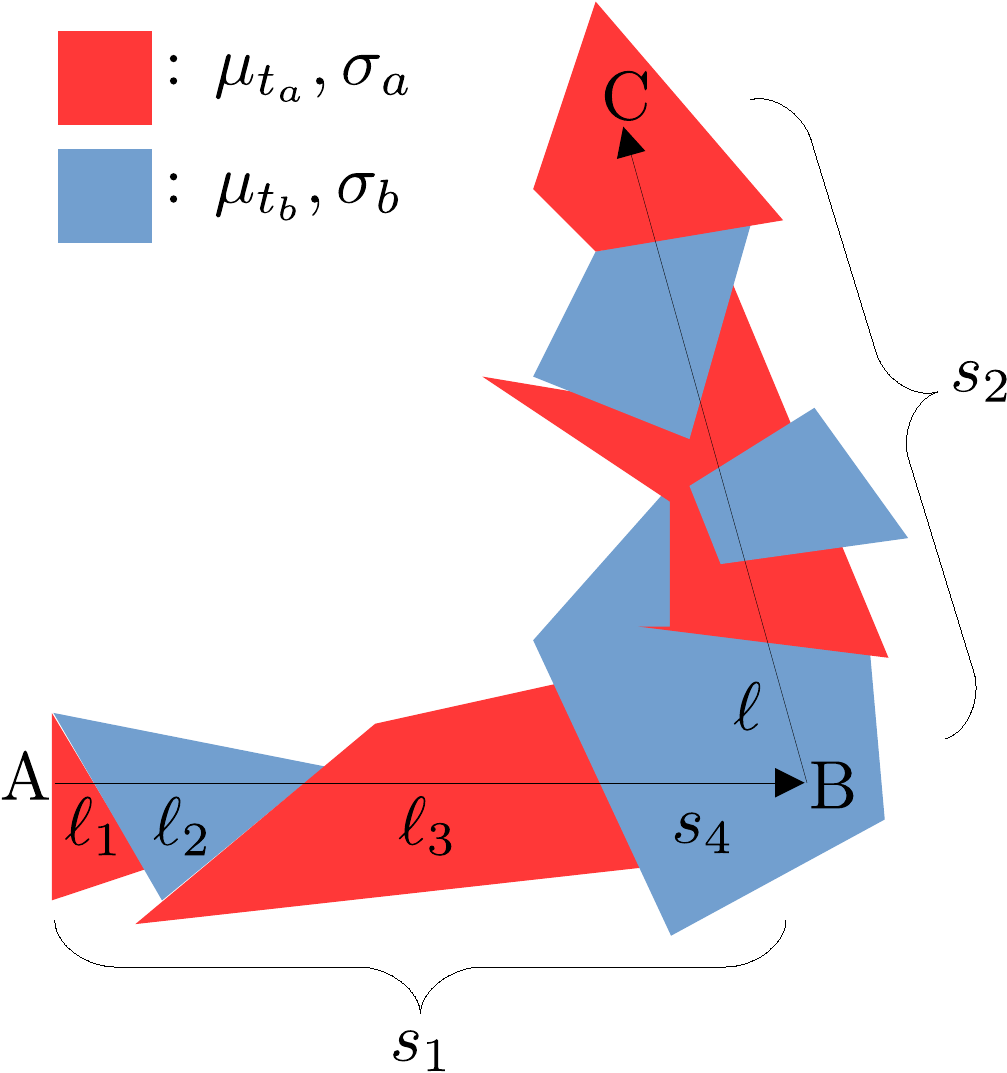}
\caption{Simplified 2D schematic representing a typical case of photon propagation in a (binary) medium.
A: Entrance point in the medium; B: The photon is scattered and changes its direction; 
C: The photon is absorbed and stops its propagation.
$s_1$: first step length; $s_2$: second step length.
In the present theoretical context and in MC simulations $s_1$ and $s_2$ are decomposed in sub-steps; e.g.,
$s_1=\ell_1+\ell_2+\ell_3+s_4$. 
Note that $s_4$ is shorter than the tessel length because
 the photon is scattered in B.
 The parameter $\ell$ is the distance from point B to the next (red) tessel.}
\label{fig:Scattering-crop}
\end{figure}

The three dimensional random medium that we will consider in the present contribution, is a medium where 
a photon that propagates in a straight line, crosses, alternatively,  two kind of homogeneous material (tessels)
(see, e.g., Fig.~\ref{fig:Scattering-crop}).
The length of each piece of path that crosses a tessel (e.g., $\ell_1, \ell_2, \ell_3$ in Fig.~\ref{fig:Scattering-crop})
is given by a probabilistic law.
If the probabilistic law remains the same for any direction of propagation of the photon, then 
the medium is called isotropic.
It has been demonstrated that if we want an isotropic medium with the above characteristics, 
then the probabilistic law must be exponential; i.e.\cite{ref:santalo2004,ref:Larmier2017},
\begin{equation}
p_{Tes}(\ell;\sigma)=\sigma e^{-\sigma \ell},
\label{eq:pdfTesselLength}
\end{equation}
where $\ell$ is the (random) length of the tessel in the direction of the photon propagation,
and the constant $\sigma \in \{\sigma_a,\sigma_b\}$, depending on the type of tessel we want to generate
(red or sky blue in Fig.~\ref{fig:Scattering-crop}).
  Observe   that the mean tessel length is 
\begin{equation}
\langle \ell \rangle=\int_0^{+\infty}p_{Tes}(\ell;\sigma)d\ell=\frac{1}{\sigma}.
\label{eq:meanpdfTesselLength}
\end{equation}
Moreover, to have an isotropic medium, $\sigma_a$ and $\sigma_b$ must satisfy the following relationship \cite{ref:santalo2004,ref:Larmier2017}
\begin{equation}
\sigma_a=\frac{1}{L}(1-
P
),
\label{eq:sigmaaL}
\end{equation}
and
\begin{equation}
\sigma_b=\frac{1}{L}
P,
\label{eq:sigmabL}
\end{equation}
where $(1-P)$ is the probability to have a tessel of type ``$a$'', and $P$ the probability to have  a tessel of type ``$b$''.
The constant $L$ is a parameter utilized in the construction of the medium.
For such a medium in the literature we speak about ``isotropic Poisson tesselation with a binary mixture''.

\subsection*{Single step function: exact analytical derivation}

In this section we will analytically derive the exact SSF, 
$p_{s_{\rm mix}}(s;\mu_{t_a},\mu_{t_b},\sigma_a,\sigma_b)$, for the 
isotropic Poisson tesselation with a binary mixture model;
where $\mu_{t_a}$ and $\mu_{t_b}$ are the extinction coefficients of the 
tessels of type ``a'' and ``b'', respectively.
The random step $s$ (ballistic propagation) is defined as a 
 straight line of random length, going  from the starting point to the point where 
the photon is absorbed or scattered. 
Thus, a step may go across many tessels (e.g., steps $s_1$ or $s_2$ in Fig.~\ref{fig:Scattering-crop}). 
By following the approach presented e.g. in Refs. \cite{ref:BinzoniMartelli2022,ref:Martelli2022} a step can be decomposed as a sum of sub-steps 
 lengths --- independently generated ---  covered by the photon in the ``a'' and ``b''   regions.
To obtain $p_{s_{\rm mix}}(.)$ we need before to derive some intermediate functions.
This is what is done in the following sub-sections.
Complex analytical calculations were performed using {\sc Mathematica}\textsuperscript{\textregistered} software.

\subsubsection*{Probability mass function  $P_N(.)$}

Probability for  a photon to make a step larger than $s$ is
\begin{equation}
P_{>s}(s;\mu_t)=\int_s^{+\infty}p_{LB}(s';\mu_t)ds'=e^{-\mu_t s}.
\end{equation}

The probability for  a photon to make a step larger than a random tessel of length $s'$ is
(the photon starts at the tessel boundary) 
\begin{align}
P_1(\mu_t,\sigma)&=\int_0^{+\infty}P_{>s}(s';\mu_t)p_{Tes}(s';\sigma)ds'
=\frac{\sigma}{\sigma+\mu_t}.
\label{eq:P1}
\end{align}
Considered the fact that the tessels crossed by a photon (ballistic propagation) have alternate  $\mu_{t_a}$ and  $\mu_{t_b}$ values,
the probability to make a series of consecutive steps larger than the correspondent random tessel lengths
must be of the form 
\begin{equation}
P_1(\mu_{t_a},\sigma_a)P_1(\mu_{t_b},\sigma_b)P_1(\mu_{t_a},\sigma_a)P_1(\mu_{t_b},\sigma_b)P_1(\mu_{t_a},\sigma_a)\dots
\label{eq:P1P1P1}
\end{equation}

Therefore, the probability $P_N(\mu_{t_a},\mu_{t_b},\sigma_a,\sigma_b,N)$ for a photon to make $N$ {\it consecutive} steps larger than the 
correspondent $N$ {\it consecutive}  random tessel lengths
(the first tessel always  has $\mu_{t_a}$) is
\begin{align}
&P_N(\mu_{t_a},\mu_{t_b},\sigma_a,\sigma_b,N)=
\nonumber \\
&\left\{
\begin{array}{lll}
1-P_1(\mu_{t_a},\sigma_a) & \mbox{if $N=0$} \\ \\
\left\{\prod_{k=1}^{N/2}\left[P_1(\mu_{t_a},\sigma_a) P_1(\mu_{t_b},\sigma_b)\right] \right\}
\left[1-P_1(\mu_{t_a},\sigma_a)\right] & \mbox{if $N$ even} \\ \\
 \left\{P_1(\mu_{t_a},\sigma_a)  
\prod_{k=1}^{(N-1)/2}\left[P_1(\mu_{t_a},\sigma_a)P_1(\mu_{t_b},\sigma_b)\right]\right\}\left[1-P_1(\mu_{t_b},\sigma_b)\right] & \mbox{if $N$ odd},
\end{array}
\right.
\label{eq:PN}
\end{align}
where $\sum_{n=0}^{+\infty}P_N(\mu_{t_a},\mu_{t_b},\sigma_a,\sigma_b,N)=1$, as expected.

In other words, if $N=0$ (first line Eq.~(\ref{eq:PN})), the photon makes a step shorter than the random length of the first tessel.
If $N$ is even, then the last term in Eq.~(\ref{eq:P1P1P1}) is $P_1(\mu_{t_b},\sigma_b)$. 
If $N$ is odd, the last term in Eq.~(\ref{eq:P1P1P1}) is $P_1(\mu_{t_a},\sigma_a)$.
This explains the terms in curly brackets in the second and third line of Eq.~(\ref{eq:PN}).
The  last photon step always has probability $1-P_1(\mu_{t_a},\sigma_a)$
or $1-P_1(\mu_{t_b},\sigma_b)$, 
$\forall N$ [Eq.~(\ref{eq:PN})], because it is where the photon stops (i.e.,
the last step must be shorter than the last random tessel length).
By inserting Eq.~(\ref{eq:P1}) in Eq.~(\ref{eq:PN}) we find
\begin{align}
&P_N(\mu_{t_a},\mu_{t_b},\sigma_a,\sigma_b,N)=
\nonumber \\
&\left\{
\begin{array}{lll}
\frac{\mu_{t_a}}{\mu_{t_a+\sigma_a}} & \mbox{if $N=0$} \\ \\
\frac{\mu_{t_a} (\sigma_a\sigma_b)^\frac{N}{2}}{(\mu_{t_a}+\sigma_a)^{\frac{N+2}{2}}(\mu_{t_b}+\sigma_b)^{\frac{N}{2}}} & \mbox{if $N$ even} \\ \\
\frac{\mu_{t_b} \sigma_a^\frac{N+1}{2} \sigma_b^\frac{N-1}{2}}{(\mu_{t_a}+\sigma_a)^{\frac{N+1}{2}}(\mu_{t_b}+\sigma_b)^{\frac{N+1}{2}}} & \mbox{if $N$ odd}.
\end{array}
\right.
\label{eq:PNfinal}
\end{align}

We derive here  the pdf $p_{s_0}(s;\mu_t,\sigma)$ for a photon step of random lengh $s$ to remain inside a tessel of random length $\ell$
(the photon always starts on the tessel boundary).
To do this, we apply an upper cut-off to Eq.~(\ref{eq:pLB}) at the distance $\ell$, i.e.,  
\begin{equation}
p_{LB_{s<\ell}}(s;\mu_t,\ell)=[1-\Theta(s-\ell)]\mu_t e^{-\mu_t s},
\end{equation}
where $\Theta(.)$ is the Heaviside function.
Thus, the  pdf  $p_{s_0}(s;\mu_t,\alpha,\sigma)$  is obtained as
\begin{equation}
p_{s_0}(s;\mu_t,\sigma)=
\frac{
\int _0^{+\infty}p_{LB_{s<\ell}}(s;\mu_t,\ell)p_{Tes}(\ell;\sigma)d\ell}{
\int _0^{+\infty}\int _0^{+\infty}p_{LB_{s<\ell}}(s;\mu_t,\ell)p_{Tes}(\ell;\sigma)d\ell ds}=
 (\mu_t+\sigma)e^{-(\mu_t+\sigma)s},
 \label{eq:p0}
\end{equation}
where the denominator appearing of Eq.~(\ref{eq:p0}) is the normalization factor.

We need also the probability $p_{Tes_N}(s;\mu_{t_a},\mu_{t_b},\sigma_a,\sigma_b,N)ds$ for a photon to make $N>1$ steps, of total random length 
$\ell_1+\ell_2+\dots+\ell_N=s\in [s-\frac{ds}{2},s+\frac{ds}{2}]$; where $\ell_i$ is the (random) length  of the $ i^{\rm th}$ tessel, in the photon direction
(the function $p_{Tes_N}(.)$ is a pdf).
To this aim, we start with the pdf $p_{Tes_1}(s;\mu_t,\sigma_t)$ for a single tessel (case $N=1$).

\subsubsection*{Probability density function $p_{Tes_N}(.)$: case $N=1$}

By applying a lower cut-off at the distance $\ell-\frac{\Delta\ell}{2}$  and an upper cut-off  at the distance $\ell+\frac{\Delta\ell}{2}$ to Eq.~(\ref{eq:pLB}) ($0<\Delta\ell \ll 1$), we express the fact  that we want the photon falls at a distance $\ell$ (tessel length in the direction of the photon propagation), i.e.,
\begin{equation}
p_{LB_{s\approx\ell}}(s;\mu_t,\ell)=\left\{\Theta\left[s-\left(\ell-\frac{\Delta\ell}{2}\right)\right]
-\Theta\left[s-\left(\ell+\frac{\Delta\ell}{2}\right)\right]\right\}\mu_t e^{-\mu_t s}.
\end{equation}
Thus
\begin{equation}
p_{Tes_1}(s;\mu_t,\sigma)\underset{N=1}{=}
\lim_{\Delta\ell \rightarrow 0}
\frac{
\int _0^{+\infty}p_{LB_{s\approx\ell}}(s;\mu_t,\ell)p_{Tes}(\ell;\sigma)d\ell}{
\int _0^{+\infty}\int _0^{+\infty}p_{LB_{s\approx\ell}}(s;\mu_t,\ell)p_{Tes}(\ell;\sigma)d\ell ds}=
(\mu_t+\sigma)e^{-(\mu_t+\sigma)s},
\label{eq:pTes11}
\end{equation}
where $\mu_t\in\{\mu_{t_a},\mu_{t_b}\}$ and where the denominator is the normalization factor allowing to obtain the pdf. Note that
\begin{equation}
p_{Tes_1}(s;\mu_t,\sigma)=p_{s_0}(s;\mu_t,\sigma).
\label{eq:ps0pDs0}
\end{equation}

Equation~(\ref{eq:pTes11}) allows one to treat the particular case of the pdf $p_{Tes_N}(s;\mu_t,\mu_t,\sigma,\sigma,N)$ of   $N$ consecutive tessels with same $\mu_t$ and $\sigma$; i.e.
\begin{equation}
p_{Tes}(s;\mu_t,\sigma,N)=p_{Tes_N}(s;\mu_t,\mu_t,\sigma,\sigma,N),
\label{eq:pTes11a}
\end{equation}
that can be explicitly written as
\begin{align}
&p_{Tes}(s;\mu_t,\sigma,N)=
\nonumber \\
&\int _0^{+\infty}\dots\int _0^{+\infty}\int _0^{+\infty}
p_{Tes_1}(\ell_1;\mu_t,\sigma) p_{Tes_1}(\ell_2;\mu_t,\sigma) \dots p_{Tes_1}(\ell_N;\mu_t,\sigma)
\delta(\ell_1+\ell_2+\dots+\ell_N-s)
d\ell_1d\ell_2\dots d\ell_N
\nonumber \\
&=
\frac{s^{N-1} (\mu_t+\sigma)^N  e^{-(\mu_t+\sigma )s}}{(N-1)!}.
\label{eq:pTes12}
\end{align}
This is the expected result since the pdf $p_{Tes}(.)$ is the sum of N positive random variables described by the 
same exponential pdf $p_{Tes_1}(.)$; resulting in a gamma function with parameters $N$ and $\mu_t+\sigma $ 
[Eq. (\ref{eq:pTes12})].
Equation~(\ref{eq:pTes12}) allows one to consider two cases for $p_{Tes_N}(s;\mu_{t_a},\mu_{t_b},\sigma_a,\sigma_b,N)$; i.e., for $N$ odd and $N$ even.

\subsubsection*{Probability density function $p_{Tes_N}(.)$: case $N>1$ odd}

If $N>1$ is odd,  by following the method proposed in Ref.~\cite{ref:Zhao2011} 
for the solution of the integral,
we obtain
\begin{align}
&p_{Tes_N}(s;\mu_{t_a},\mu_{t_b},\sigma_a,\sigma_b,N)= \nonumber \\
&\int_0^{+\infty}\int_0^{+\infty}p_{Tes}(\ell_1;\mu_{t_a},\sigma_a,\frac{N+1}{2})p_{Tes}(\ell_2;\mu_{t_b},\sigma_b,\frac{N-1}{2})
\delta(\ell_1+\ell_2-s)d\ell_1d\ell_2 
\nonumber \\
&=p_k(s;\mu_{t_a},\mu_{t_b},\sigma_a,\sigma_b,N)\int_0^1 p_g(w,s;\mu_{t_a},\mu_{t_b},\sigma_a,\sigma_b,N)dw,
\label{eq:pTesN1}
\end{align}
where
\begin{equation}
p_k(s;\mu_{t_a},\mu_{t_b},\sigma_a,\sigma_b,N)=
\frac{(\mu_{t_b}+\sigma_b)^{\frac{N-1}{2}}(\mu_{t_a}+\sigma_a)^{\frac{N+1}{2}}}{2^{N-1}\Gamma(\frac{N-1}{2})\Gamma(\frac{N+1}{2})}
s^{N-1}e^{-\frac{(\mu_{t_a}+\mu_{t_b}+\sigma_a+\sigma_b)}{2}s},
\end{equation}
and
\begin{equation}
p_g(w,s;\mu_{t_a},\mu_{t_b},\sigma_a,\sigma_b,N)= 
\left(1-w^2\right)^{\frac{N-3}{2}}
\left[\left(1-w\right)e^{-\frac{\mu_{t_b}-\mu_{t_a}+\sigma_b-\sigma_a}{2}s w}+
\left(1+w\right)e^{\frac{\mu_{t_b}-\mu_{t_a}+\sigma_b-\sigma_a}{2}s w}\right],
\end{equation}
and
\begin{align}
\int_0^1 p_g(w,s;\mu_{t_a},\mu_{t_b},\sigma_a,\sigma_b,N)dw= 
\frac{\sqrt{\pi }\,\Gamma(\frac{N-1}{2})}{2^{N-2}\mu^{\frac{N-1}{2}}}
\left[I_{\frac{N-2}{2}}\left(\frac{s \mu }{2}\right)+I_{\frac{N}{2}}\left(\frac{s \mu }{2}\right)\right]
\frac{1}{s^{\frac{N-1}{2}}}
\end{align}
where 
\begin{equation}
\mu=\mu_{t_b}-\mu_{t_a} +\sigma_b-\sigma_a.
\end{equation}
Thus,
\begin{align}
 p_{Tes_N}(s;\mu_{t_a},\mu_{t_b},\sigma_a,\sigma_b,N)
\underset{\begin{subarray}{c} N>1 \\ N \mathrm{\phantom{a}odd}\end{subarray} }{=}
  \frac{\sqrt{\pi }   (\mu_{t_b}+\sigma_b)^{\frac{N-1}{2}}(\mu_{t_a}+\sigma_a)^{\frac{N+1}{2}}   }{2\mu^{\frac{N-2}{2}} 
  \Gamma \left(\frac{N+1}{2}\right)}
s^{\frac{N}{2}} 
\left[I_{\frac{N-2}{2}}\left(\frac{s \mu }{2}\right)+I_{\frac{N}{2}}\left(\frac{s \mu }{2}\right)\right]
  e^{-\frac{(\mu_{t_a}+\mu_{t_b}+\sigma_a+\sigma_b)}{2}s} \, .
\label{eq:pTesOddCompact}
\end{align}
where $I_{n}(x)$ is the modified Bessel function of the first kind.


Note that the method proposed in Ref.~\cite{ref:Zhao2011} has been developed by imposing some constraints on the parameters $\mu_{t_a}$, $\mu_{t_b}$,
$\sigma_a$ and $\sigma_b$.
$\sigma_a$ and $\sigma_b$ value.
 However, it is easy to show that the method remains valid for any value of $\mu_{t_a}$, $\mu_{t_b}$,
$\sigma_a$ and $\sigma_b$. 

\subsubsection*{Probability density function $p_{Tes_N}(.)$: case $N>1$ even} 

If $N>1$ is even,   in the same vein,  by following the method proposed in Ref. \cite{ref:Zhao2011} 
for the solution of the integral,
we get
\begin{align}
&p_{Tes_N}(s;\mu_{t_a},\mu_{t_b},\sigma_a,\sigma_b,N)= \nonumber \\
&\int_0^{+\infty}\int_0^{+\infty}p_{Tes}(\ell_1;\mu_{t_a},\sigma_a,\frac{N}{2})p_{Tes}(\ell_2;\mu_{t_b},\sigma_b,\frac{N}{2})
\delta(\ell_1+\ell_2-s)d\ell_1d\ell_2 \nonumber \\
&=p_k(s;\mu_{t_a},\mu_{t_b},\sigma_a,\sigma_b,N)\int_0^1 p_g(w,s;\mu_{t_a},\mu_{t_b},\sigma_a,\sigma_b,N)dw,
\label{eq:pTesN}
\end{align}
where
\begin{equation}
p_k(s;\mu_{t_a},\mu_{t_b},\sigma_a,\sigma_b,N)=
\frac{(\mu_{t_b}+\sigma_b)^{\frac{N}{2}}(\mu_{t_a}+\sigma_a)^{\frac{N}{2}}}{2^{N-1}
\Gamma(\frac{N}{2})^2 }
s^{N-1}e^{-\frac{(\mu_{t_a}+\mu_{t_b}+\sigma_a+\sigma_b)}{2}s},
\end{equation}
and
\begin{equation}
p_g(w,s;\mu_{t_a},\mu_{t_b},\sigma_a,\sigma_b,N)= 
\left(1-w^2\right)^{\frac{N-2}{2}}
\left[e^{-\frac{\mu_{t_b}-\mu_{t_a}+\sigma_b-\sigma_a}{2}s w}+
e^{\frac{\mu_{t_b}-\mu_{t_a}+\sigma_b-\sigma_a}{2}s w}\right],
\end{equation}
and
\begin{align}
\int_0^1 p_g(w,s;\mu_{t_a},\mu_{t_b},\sigma_a,\sigma_b,N)dw=
 \frac{\sqrt{\pi}\,\Gamma(\frac{N}{2})}{2^{N-1} \mu^{\frac{N-1}{2}} }
I_{\frac{N-1}{2}}\left(\frac{s \mu }{2}\right)
\frac{1}{s^{\frac{N-1}{2}}}
\end{align}
where 
\begin{equation}
\mu=\mu_{t_b}-\mu_{t_a} +\sigma_b-\sigma_a.
\end{equation}
Thus,
 \begin{equation}
p_{Tes_N}(s;\mu_{t_a},\mu_{t_b},\sigma_a,\sigma_b,N) \underset{\begin{subarray}{c} N>1 \\ N \mathrm{\phantom{a}even}\end{subarray} }{=} 
  \frac{\sqrt{\pi }   (\mu_{t_b}+\sigma_b)^{\frac{N}{2}}(\mu_{t_a}+\sigma_a)^{\frac{N}{2}}   }{\mu^{\frac{N-1}{2}} 
  \Gamma \left(\frac{N}{2}\right)}
s^{\frac{N-1}{2}} I_{\frac{N-1}{2}}\left(\frac{s \mu }{2}\right)  e^{-\frac{(\mu_{t_a}+\mu_{t_b}+\sigma_a+\sigma_b)}{2}s} \, .
\label{eq:pTesNevenCompact}  
\end{equation}
Note that, in general,  $\int_0^{+\infty} p_{Tes_N}(s;\mu_{t_a},\mu_{t_b},\sigma_a,\sigma_b,N) ds =1, \forall N\ge 1$.
\subsubsection*{Probability density function $p_{s_{\rm mix}}(s;.)$ (single step function)} 

Let's first express the pdf $p_{s_N}(s;\mu_{t_a},\mu_{t_b},\sigma_a,\sigma_b,N)$  for a photon to jump over $N$ tessels and reach a distance s.
In practice, the photon stops inside the $(N+1)^{\rm th}$ tessel due to absorption or scattering.
See, e.g., an intuitive drawing in Fig. \ref{fig:Scattering-crop} for the scattering case 
(segment AB where $N=3$).
This is obtained  as  [Eqs. (\ref{eq:ps0pDs0}), (\ref{eq:pTesN1}) and (\ref{eq:pTesN})]
\begin{align}
&p_{s_N}(s;\mu_{t_a},\mu_{t_b},\sigma_a,\sigma_b,N)= \nonumber \\
&\int_0^{+\infty }\int_0^{+\infty }p_{Tes_N}(s_1;\mu_{t_a},\mu_{t_b},\sigma_a,\sigma_b,N)p_{s_0}(s_2;\mu_t,\sigma_t)\delta(s_1+s_2-s)ds_1ds_2 \nonumber \\
&=\int_0^{+\infty }\int_0^{+\infty }p_{Tes_N}(s_1;\mu_{t_a},\mu_{t_b},\sigma_a,\sigma_b,N)p_{Tes_1}(s_2;\mu_t,\sigma_t)\delta(s_1+s_2-s)ds_1ds_2 \nonumber \\
&=p_{Tes_N}(s;\mu_{t_a},\mu_{t_b},\sigma_a,\sigma_b,N+1),
\label{eq:psNpsN}
\end{align}
where $(\mu_t=\mu_{t_a}  \land \sigma_t=\sigma_a)$ or 
$(\mu_t=\mu_{t_b}  \land \sigma_t=\sigma_b)$, depending if $N$ is even or odd, respectively. 

Thus, the pdf $p_{s}(s;\mu_{t_a},\mu_{t_b},\sigma_a,\sigma_b)$ for a photon to make a step of length $s$,
independently of the number of tessels $N$, is obtained as [Eqs. (\ref{eq:PNfinal}) and (\ref{eq:psNpsN})]
\begin{align}
p_{s}(s;\mu_{t_a},\mu_{t_b},\sigma_a,\sigma_b)&=
\sum_{N=0}^{+\infty}p_{s_N}(s;\mu_{t_a},\mu_{t_b},\sigma_a,\sigma_b,N)
P_N(\mu_{t_a},\mu_{t_b},\sigma_a,\sigma_b,N)
\nonumber \\
&=
\sum_{N=0}^{+\infty}p_{Tes_N}(s;\mu_{t_a},\mu_{t_b},\sigma_a,\sigma_b,N+1)
P_N(\mu_{t_a},\mu_{t_b},\sigma_a,\sigma_b,N).
\label{eq:psFINAL}
\end{align}
Equation (\ref{eq:psFINAL}) has been derived  for a starting tessel with parameters  $\mu_{t_a}$ and 
$\sigma_a$.
However, Eq.~(\ref{eq:psFINAL}) also works  if we permute  $\mu_{t_a}$ with $\mu_{t_b}$, and  $\sigma_a$
with $\sigma_b$ in the equation (i.e., we chose the parameters  of the starting tessel  equal to $\mu_{t_b}$
and $\sigma_b$)

Finally, Eq.~(\ref{eq:psFINAL}) allows us to express in general the pdf 
$p_{s_{\rm mix}}(s;\mu_{t_a},\mu_{t_b},\sigma_a,\sigma_b)$ (single step function)  for a photon step of length $s$, for a medium
with a probability $1-P$ to have the first tessel with parameters $\mu_{t_a}$ and $\sigma_a$,
and probability $P$ to have the first tessel with parameters $\mu_{t_b}$ and $\sigma_b$; i.e.,
\begin{equation}
p_{s_{\rm mix}}(s;\mu_{t_a},\mu_{t_b},\sigma_a,\sigma_b,
P
)=
(1-P) p_{s}(s;\mu_{t_a},\mu_{t_b},\sigma_a,\sigma_b)+P p_{s}(s;\mu_{t_b},\mu_{t_a},\sigma_b,\sigma_a).
\label{eq:psFINALmix}
\end{equation}
Equation (\ref{eq:psFINALmix}) has been implemented in {\sc Matlab}\textsuperscript{\textregistered} language. 
Note that, to obtain Eq.~(\ref{eq:psFINALmix}), we did not use the isotropy conditions
[Eqs.~(\ref{eq:sigmaaL}) and (\ref{eq:sigmabL})],
thus the pdf $p_{s_{\rm mix}}(.)$ remains valid 
even if the parameters change with the propagation direction.

\subsection*{Derivation of the albedo for the random medium with mixing statistics}

The albedo for a homogeneous medium is usually expressed as $\mu_s/\mu_t$,
where $\mu_s$ is the scattering coefficient, but where, in probabilistic terms, $\mu_s  ds$ is interpreted
as the probability to be scattered inside a length $ds$.
Thus,  $\mu_s$ can be seen as the probability of scattering per unit length. 
In the same way, $\mu_t=\mu_a+\mu_s$ may be interpreted
as the probability for a photon to be scattered or absorbed (extinction) per unit length;
where $\mu_a$ is the absorption coefficient.
In the present case, the ``probabilities'' $\mu_s$ and $\mu_t$ must be re-calculated,
to take into account of the effects of the mixing statistics.
To this aim, the probability, $P_{t_{unit}}\Delta s$,  to occur in an extinction interaction, inside a very small $\Delta s$, is calculated as:
\begin{equation}
P_{t_{\rm unit}}\Delta s=\int_0^{\Delta s} p_{s_{\rm mix}}(s';\mu_{t_a},\mu_{t_b},\sigma_a,\sigma_b,
P
)ds
\approx
[(1-P)\mu_{t_a}+P\mu_{t_b}]\Delta s,
\end{equation}
where we have developed in Taylor series   the integral  for $\Delta s \ll 1$.
Equivalently, the probability,  $P_{s_{unit}}\Delta s$,  for a photon to occur in a scattering event in $\Delta s$ is
expressed as 
\begin{equation}
P_{s_{\rm unit}}\Delta s=\int_0^s p_{s_{\rm mix}}(s';\mu_{s_a},\mu_{s_b},\sigma_a,\sigma_b,
P
)ds
\approx
[(1-P)\mu_{s_a}+P\mu_{s_b}]\Delta s,
\end{equation}
where we have set $\mu_{a_a}=\mu_{a_b}=0$.
Finally, the albedo $\Lambda_{\rm mix}$, to be utilized with the SSF $p_{s_{\rm mix}}(.)$, is obtained as
\begin{equation}
\Lambda_{\rm mix}=\frac{P_{s_{\rm unit}}}{P_{t_{\rm unit}}}=\frac{(1-P)\mu_{s_a}+P\mu_{s_b}}{(1-P)\mu_{t_a}+P\mu_{t_b}}.
\end{equation}
Note that if $\mu_{t_a}=\mu_{t_b}$ and $\mu_{s_a}=\mu_{s_b}$, we obtain the classical 
albedo for homogeneous media.

\subsection*{Derivation of the single step function for photons starting from ``fixed'' positions}

For the sake of completeness, in this section we give the essential equations allowing the reader to derive 
the SSP of photons starting from fixed positions (this topic may be subject of further studies).
It has been shown that when $p_{s_{\rm mix}}(.)$ [Eq. (\ref{eq:psFINALmix})] is not a pure exponential law,
photon steps starting from fixed positions (e.g., from a fixed light source or after the reflection on a medium boudary) must 
satisfy the following law \cite{ref:dEon2018,ref:BinzoniMartelli2022}
\begin{align}
p_{s_{\rm fix}}(s;\mu_{t_a},\mu_{t_b},\sigma_a,\sigma_b)=\frac{1-\int_0^{s} p_{s_{\rm mix}}(s';\mu_{t_a},\mu_{t_b},\sigma_a,\sigma_b)ds'}
{\int_0^{+\infty} s' p_{s_{\rm mix}}(s';\mu_{t_a},\mu_{t_b},\sigma_a,\sigma_b)ds'},  
\label{eq:uncorr}
\end{align}
where $p_{s_{\rm fix}}(.)$ is always a pdf.
Using Eq. (\ref{eq:psFINAL}) and (\ref{eq:psFINALmix}), Eq. (\ref{eq:uncorr}) can be written as
\begin{align}
&p_{s_{\rm fix}}(s;\mu_{t_a},\mu_{t_b},\sigma_a,\sigma_b)=\frac{1-\int_0^{s} p_{s_{\rm mix}}(s';\mu_{t_a},\mu_{t_b},\sigma_a,\sigma_b)ds'}
{\int_0^{+\infty} s' p_{s_{\rm mix}}(s';\mu_{t_a},\mu_{t_b},\sigma_a,\sigma_b)ds'} 
\nonumber \\&=
\frac{1- (1-P) \sum_{N=0}^{+\infty}[F(s;a,b,N+1)
P_N(\mu_{t_a},\mu_{t_b},\sigma_a,\sigma_b,N)]+
P \sum_{N=0}^{+\infty}[F(s;b,a,N+1)
P_N(\mu_{t_b},\mu_{t_a},\sigma_b,\sigma_a,N)]}
{(1-P)\frac{\mu_{t_b}+\sigma_a+\sigma_b}{\mu_{t_b}\sigma_a+\mu_{t_a}(\mu_{t_b}+\sigma_b)} +
P \frac{\mu_{t_a}+\sigma_b+\sigma_a}{\mu_{t_a}\sigma_b+\mu_{t_b}(\mu_{t_a}+\sigma_a)} },
\label{eq:uncorr2}
\end{align}
where
\begin{align}
F(s;a,b,N)=\int_0^{s}p_{Tes_N}(s';\mu_{t_a},\mu_{t_b},\sigma_a,\sigma_b,N)ds'
\label{eq:F}
\end{align}
Note that the denominator of Eq. (\ref{eq:uncorr2}) is the mean photon step length $\langle s \rangle_{\rm mix}$ in the binary statistical mixture (if $\mu_{t_a}=\mu_{t_b}$ we obtain $\langle s \rangle_{\rm mix}=\frac{1}{\mu_{t_a}}$, as expected).
The function $F(s;a,b,N)$ can be expressed as (see the appendix for calculation details)
\begin{align}
F(s;a,b,N) \underset{\begin{subarray}{c} N=1 \end{subarray} }{=} 
1-e^{-(\mu_{t_a}+\sigma_a)s},
\label{eq:F1}
\end{align}
\begin{align}
&F(s;a,b,N) \underset{\begin{subarray}{c} N>1 \\ N \mathrm{\phantom{a}even}\end{subarray} }{=} 
 \nonumber \\ & 1 +
e^{-(\mu_{t_a}+\sigma_a) s}\sum_{k=1}^{N/2}\frac{s^{k-1}}{\Gamma(k)}
\left[-(\mu_{t_a}+\sigma_a)^{k-1}\right.
\nonumber \\  &\phantom{1} +
\left.
(-1)^{\frac{N+2}{2}}(-\mu)^{k-N}
(\mu_{t_a}+\sigma_a)^{\frac{N}{2}} (\mu_{t_b}+\sigma_b)^{\frac{N-2}{2}}\binom{N-k}{\frac{N}{2}-k+1} \, _2F_1\left(1,1-\frac{N}{2};\frac{N}{2}-k+2;\frac{\mu_{t_a}+\sigma_a}{\mu_{t_b}+\sigma_b}\right)\right]
\nonumber \\ &\phantom{1} +
e^{-(\mu_{t_b}+\sigma_b) s}\sum_{k=1}^{N/2}\frac{s^{k-1}}{\Gamma(k)}
\left[-(\mu_{t_b}+\sigma_b)^{k-1}\right.
\nonumber \\  &\phantom{1} +
\left.
(-1)^{\frac{N+2}{2}}(\phantom{-}\mu)^{k-N}
(\mu_{t_b}+\sigma_b)^{\frac{N}{2}} (\mu_{t_a}+\sigma_a)^{\frac{N-2}{2}}\binom{N-k}{\frac{N}{2}-k+1} \, _2F_1\left(1,1-\frac{N}{2};\frac{N}{2}-k+2;\frac{\mu_{t_b}+\sigma_b}{\mu_{t_a}+\sigma_a}\right)\right],
\label{eq:Feven}
\end{align}
where $\binom{.}{.}$ is the binomial coefficient and $_2F_1(.,.;.;.)$ the hypergeometric function;
and
\begin{align}
&F(s;a,b,N) \underset{\begin{subarray}{c} N>1 \\ N \mathrm{\phantom{a}odd}\end{subarray} }{=}
\nonumber \\ &1  + 
e^{-(\mu_{t_a}+\sigma_a) s} (\mu_{t_b}+\sigma_b)^\frac{N-1}{2} \sum _{k=0}^{\frac{N-1}{2}} \frac{s^{k}}{\Gamma (k+1)} \frac{1}{ (-\mu)^{N-1-k}}  \nonumber \\ &\phantom{1} \times
\left[ (-1)^{\frac{N+1}{2}} (\mu_{t_a}+\sigma_a)^{\frac{N-1}{2}} 
\binom{N-1-k}{\frac{N-1}{2}-k} \, _2F_1\left(1,k-\frac{N-1}{2};\frac{N+1}{2} ;\frac{\mu_{t_b}+\sigma_b}{\mu_{t_a}+\sigma_a}\right) \right.\nonumber \\
&\phantom{1}\left. - \frac{(\mu_{t_b}+\sigma_b)^{\frac{N+1}{2}}}{\mu_{t_a}+\sigma_a} \binom{N-1-k}{-k-1} \, _2F_1\left(1,k+1;  N+1;\frac{\mu_{t_b}+\sigma_b}{\mu_{t_a}+\sigma_a}\right)\right] \nonumber \\
&\phantom{1}+ e^{-(\mu_{t_b}+\sigma_b) s} (\mu_{t_a}+\sigma_a)^{\frac{N+1}{2}}  \sum _{k=0}^{\frac{N-1}{2}} \frac{s^{k}}{\Gamma (k+1)} \frac{1}{ (-\mu)^{N-1-k}}  
\nonumber \\ &\phantom{1} \times
\left[ (-1)^{k+\frac{N-1}{2}} (\mu_{t_b}+\sigma_b)^{\frac{N-3}{2}} \binom{N-1-k}{\frac{N-3}{2}-k} \, _2F_1\left(1,k-\frac{ N-3}{2};\frac{N+3}{2};\frac{\mu_{t_a}+\sigma_a}{\mu_{t_b}+\sigma_b}\right)  \right.  \nonumber \\
 &\phantom{1}\left.  + (-1)^{k+1} \frac{(\mu_{t_a}+\sigma_a)^\frac{N-1}{2}}{\mu_{t_b}+\sigma_b} \binom{N-1-k}{-k-1} \, _2F_1\left(1,k+1;N+1;\frac{\mu_{t_a}+\sigma_a}{\mu_{t_b}+\sigma_b}\right) \right]  \nonumber \\ 
\label{eq:Fodd}
\end{align}
In MC simulations, the use of $p_{s_{\rm fix}}(.)$ is fundamental,  because it allows one to preserve the invariance property for light 
propagation and the related reciprocity law.\cite{ref:BinzoniMartelli2022}

\subsection*{Single step function: direct Monte Carlo simulation}

\subsubsection*{Numerical generation of Eq.~(\ref{eq:psFINALmix})}

Equation (\ref{eq:psFINALmix}) has been validated numerically by MC simulation; i.e., 
by explicitly taking into account each single tessel crossed by the photons.
Considering that by definition $p_{s_{\rm mix}}(s;\mu_{t_a},\mu_{t_b},\sigma_a,\sigma_b)$ 
takes into account only ``ballistic'' photons (until they are absorbed or scattered),
the code has been implemented --- in {\sc Matlab}\textsuperscript{\textregistered} language --- 
in the following manner:
\begin{enumerate}[label=\arabic*)]
  \item A uniformly distributed random number $\xi\in\{0,1\}$ is generated;
  \item If $\xi<(1-P)$ then the parameters used   are $\mu_{t_a}$ and $\sigma_a$, 
otherwise $\mu_{t_b}$ and $\sigma_b$;
  \item A random tessel of length $\ell$  is generated by means of Eq.~(\ref{eq:pdfTesselLength})
and the chosen $\sigma_{t_i}$ of point 2 ($i\in\{a,b\}$); 
  \item A random photon step $s$ is generated by means of Eq.~(\ref{eq:pLB})
and the chosen $\mu_{t_i}$ of point 2 ($i\in\{a,b\}$);
  \item If $s>l$ then permute the parameters (i.e., permute a and b parameters) and go to 3,
otherwise stop the photon (because absorbed or scattered inside this last tessel) and store the total length traveled and the number of tessels crossed by the photon;
  \item Go to 1 until the desired number of photons are propagated.
\end{enumerate}

Note that, usually, in complex MC simulations many photons are launched for a given random tessel configuration.
Then, this procedure is repeated a number of times and  the ``ensemble'' average of the results is taken.
This approach is imposed by the intensive computation demand of the MC code.
However, the same results can be obtained by changing tessel configuration for each launched photon.
This is what is done in the present simpler MC context (points 1 to 6 above).

\subsubsection*{Single step function for photons starting at any point inside a tessel}

Equation~(\ref{eq:psFINALmix}) has been derived for photons starting at the medium boundary,
and thus also from any tessel boundary (see, e.g., point A in Fig.~\ref{fig:Scattering-crop}).
However, in real MC simulations  photons propagate through the medium
and  steps may start at any point inside a tessel (see, e.g., point B in Fig. \ref{fig:Scattering-crop}).
In the present context this is not a problem, because Eq.~(\ref{eq:pdfTesselLength})  is a memoryless law (exponential pdf),
and thus starting from a tessel boundary or inside the tessel  does not change the present findings.

To give a more intuitive view on this point, some tutorial MC simulations  have been performed.
These simulations have the aim to show that Eq.~(\ref{eq:psFINALmix})
remains valid even in the case   where  the photons start from points situated inside the tessels.
(see, e.g., the $s_2$ step in Fig~\ref{fig:Scattering-crop}).
Considered that the medium is isotropic, the MC code has been implemented in the following manner
(see also Fig.~\ref{fig:StartingPoint-crop}):
\begin{figure}[ht]
\centering
\includegraphics[width=0.55\linewidth]{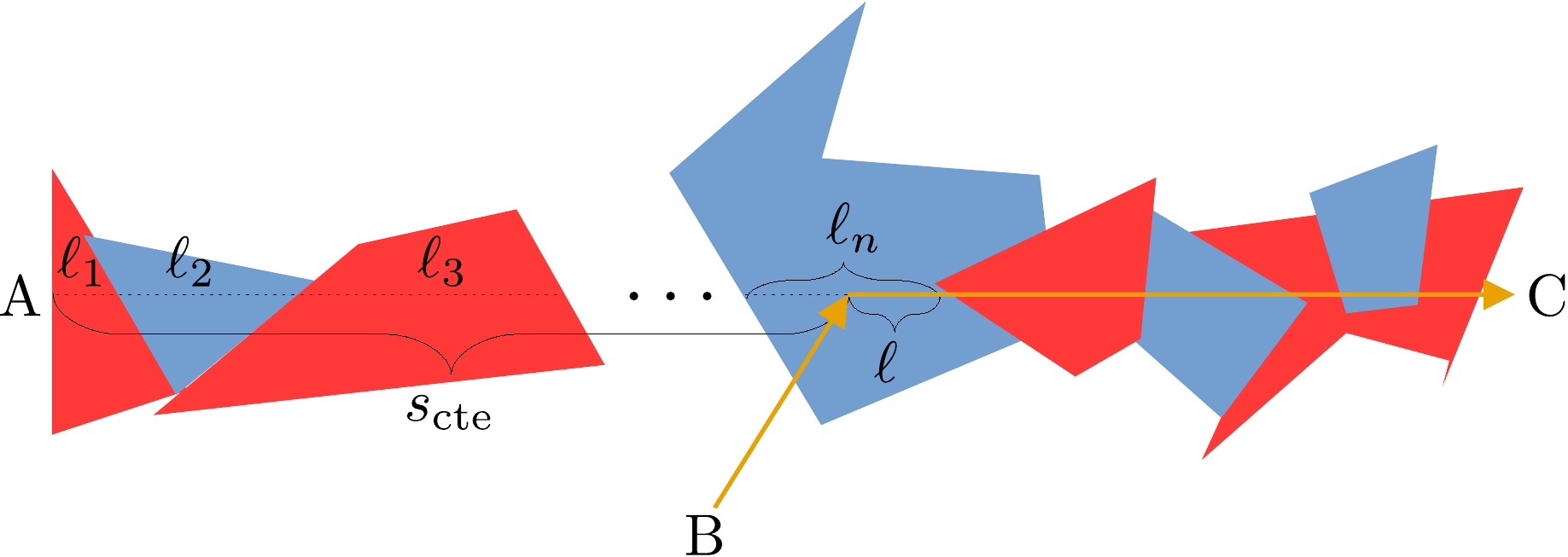}
\caption{A photon coming from B scatters and goes in direction C.
The scattering point is at distance $s_{\rm cte}$ from the boundary, measured along 
the scattered direction.}
\label{fig:StartingPoint-crop}
\end{figure}

\begin{enumerate}[label=\arabic*)]
  \item A random $s_{\rm cte} \in [0,+\infty]$, representing the distance from any point on the  medium boundary (e.g., point A in Fig.~\ref{fig:StartingPoint-crop}) along the considered direction of photon propagation, is chosen; 
  \item Random tessel lengths $\ell_1, \ell_2, \dots$  are generated using Eq.~(\ref{eq:pdfTesselLength}),
 by alternatively using $\sigma=\sigma_a$ and $\sigma=\sigma_b$, until for a given $n$, $\ell_1 + \ell_2+ \dots \ell_n > s_{\rm cte}$;
  \item $\ell=s_{\rm cte}-(\ell_1 + \ell_2+ \dots \ell_n)$ is computed;
  \item Points 2 and 3 are repeated many times and the different $\ell$ are saved;
  \item The pdf for $\ell$ is obtained by computing the histogram of the $\ell $ data saved in point 4.
\end{enumerate}
   
 In practice, we want to demonstrate that the pdf for  $\ell_n$ (see, Fig.~\ref{fig:StartingPoint-crop})
 is the same as the pdf for $\ell$; i.e., we always have Eq.~(\ref{eq:pdfTesselLength});
 where $n\in\{1,2,\dots\}$ represents any $n^{\rm th}$ tessel where the photon is extinct.
This means that, it is not important if the photon starts from the tessel boundary or inside the tessel,
 because Eq.~(\ref{eq:psFINALmix}) remains the same (i.e., the previous mathematical derivation
 of  Eq.~(\ref{eq:psFINALmix})  does not change).

\section*{Explanatory examples}

In the following subsections we will show the validity of Eq.~(\ref{eq:psFINALmix})
through some intuitive explanatory examples, and comparisons with 
MC simulations. 

\subsection*{Example 1}

One of the simplest cases we can study is when one of the tessels type (e.g., ``b'')
has mean length zero [Eq.~(\ref{eq:meanpdfTesselLength})].
This means that we remain with an homogeneous medium of type ``a'' (only sky blue tessels).
In this case, we must retrieve the classical exponential law [Eq.~(\ref{eq:pLB})].
  Indeed,
\begin{equation}
\lim_{\sigma_b \rightarrow +\infty} p_{s_{\rm mix}}(s;\mu_{t_a},\mu_{t_b},
\sigma_a,\sigma_b,
P
)=
\mu_{t_a}e^{-\mu_{t_a} s},
\end{equation}
as it is expected.
In the case of an isotropic medium, also $\sigma_a$ must go to $+\infty$ because this quantity is linked to
$\sigma_b$ by Eqs.~(\ref{eq:sigmaaL}) and (\ref{eq:sigmabL}), i.e.; $L\rightarrow 0$.
In this latter particular case $p_{s_{\rm mix}}(.)$=0, because the mean length of the tessels is zero,
and thus the propagating medium does not exist.

\subsection*{Example 2}

Another simple case is when the mean lengths [Eq.~(\ref{eq:meanpdfTesselLength})] of both tessel  types
are infinite.
In this case, we expect that the photon stops inside the first tessel at the entrance of the medium.
Considering that tessel types ``a'' and ``b'' have probability $P$ and $(1-P)$
to be the first tessel, we also expect  to obtain the sum of two
 classical exponential laws [Eq.~(\ref{eq:pLB})]
weighted by their probability.
In fact, 
\begin{equation}
\lim_{\substack{\sigma_a \rightarrow 0 \\ \sigma_b \rightarrow 0}} 
p_{s_{\rm mix}}(s;\mu_{t_a},\mu_{t_b},\sigma_a,\sigma_b,
P
)=
(1-P)\mu_{t_a}e^{-\mu_{t_a} s}+P\mu_{t_b}e^{-\mu_{t_b} s},
\label{eq:Ex2}
\end{equation}
as it must be.
If the medium is isotropic
[Eqs.~(\ref{eq:sigmaaL}) and (\ref{eq:sigmabL}); $L\rightarrow +\infty$]
the result remains the same.

Due to the fact that Eq. (\ref{eq:Ex2}) is not a pure exponential law, 
photon steps that start from a fixed position must be described using Eq. (\ref{eq:uncorr}), i.e.;
\begin{equation}
\lim_{\substack{\sigma_a \rightarrow 0 \\ \sigma_b \rightarrow 0}}p_{s_{\rm fix}}(s;\mu_{t_a},\mu_{t_b},\sigma_a,\sigma_b)=
\frac{\mu_{t_a}\mu_{t_b}}{(1-P)\mu_{t_b}+P\mu_{t_a}}\left[(1-P)\mu_{t_a}e^{-\mu_{t_a} s}+P\mu_{t_b}e^{-\mu_{t_b} s}\right],
\end{equation}

\subsection*{Example 3}

In the case $\mu_{t_a}=\mu_{t_b}$ and $\sigma_a=\sigma_b $ 
[see, Eqs.~(\ref{eq:sigmaaL}) and (\ref{eq:sigmabL})],
we are in the presence
of a homogeneous medium, and it is possible to obtain the explicit solution for 
$p_{s_{\rm mix}}(.)$, as 
\begin{equation}
p_{s_{\rm mix}}(s;\mu_{t_a},\mu_{t_a},\sigma_a,\sigma_a,P)=
\mu_{t_a}e^{-\mu_{t_a} s}; \quad 0 \le P \le 1,
\label{eq:ex3}
\end{equation}
where Eq.~(\ref{eq:ex3}) represents the expected pdf for the homogeneous medium
with extinction coefficient $\mu_{t_a}$.
The parameter $\sigma_a$ has disappeared because it is no more possible to distinguish the tessels, i.e.;
they all have the same optical parameters.
It obviously follows, that if the medium is isotropic --- i.e.; for  $\mu_{t_a}=\mu_{t_b}$ and $\sigma_a=\sigma_b =\frac{1}{2 L}$ 
[see, Eqs.~(\ref{eq:sigmaaL}) and (\ref{eq:sigmabL})], when $P=\frac{1}{2}$ ---
Eq.~(\ref{eq:ex3}) remains valid.

\subsection*{Example 4}

\begin{figure}[ht]
\centering
\includegraphics[width=0.4\linewidth]{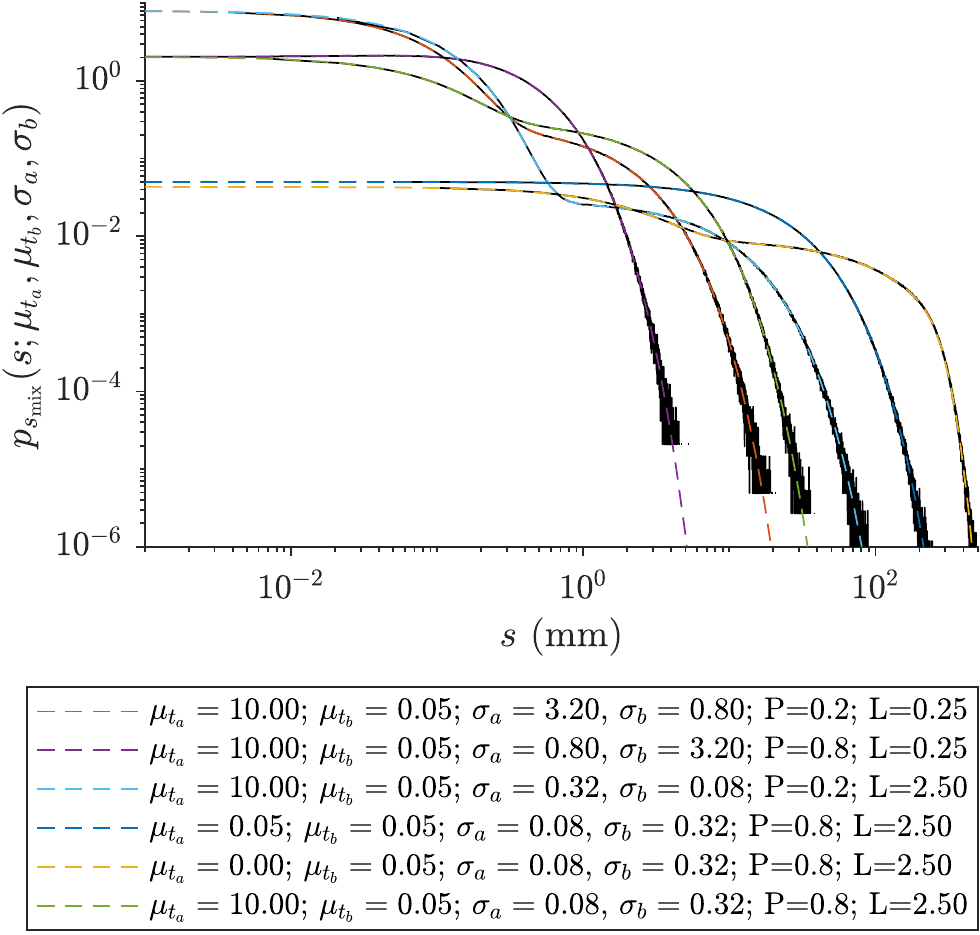}
\caption{The SSF $p_{s_{\rm mix}}(.)$ as a function of $s$ [Eq.~(\ref{eq:psFINALmix})] for a medium with isotropic Poisson tesselation and binary mixture, and for a set different optical and geometrical parameters. 
Black lines represent the relative  MC data.}
\label{fig:FigPaper1-crop}
\end{figure}

To investigate more complex cases, we need to compare our analytical model 
for the SSF [Eq.~(\ref{eq:psFINALmix})]
with the relative ``gold standard'' MC simulations.
Figure~\ref{fig:FigPaper1-crop} 
shows the SSF $p_{s_{\rm mix}}(s;.)$  for different optical and geometrical values,
compared to the  MC simulations.
It can be seen that the analytical method gives the same results as the reference MC data.

\subsection*{Example 5}

\begin{figure}[ht]
\centering
\includegraphics[width=0.5\linewidth]{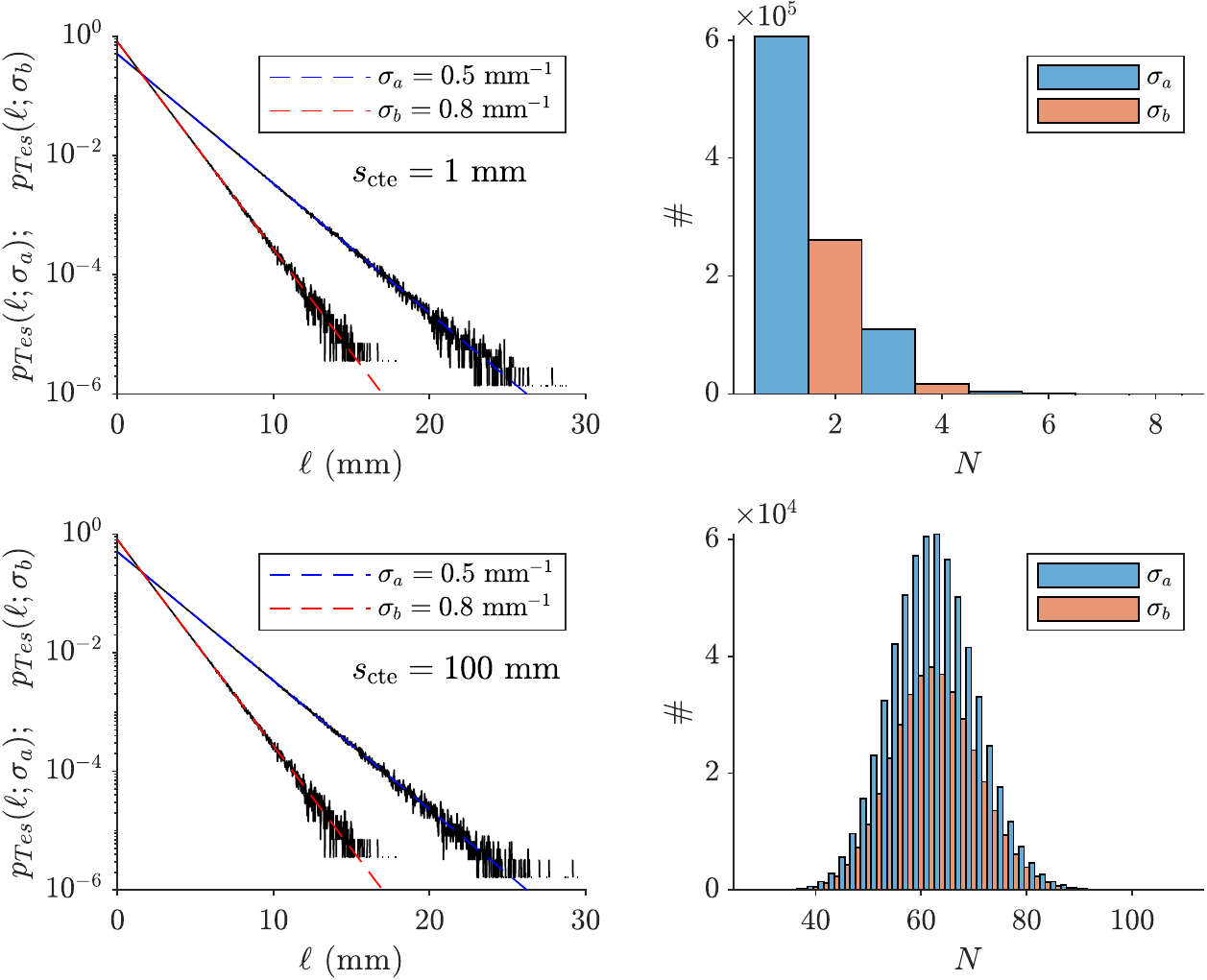}
\caption{Dashed lines represent Eq.~(\ref{eq:pdfTesselLength}).
Black lines are MC simulations.
The symbol \# represents  the number of tessels $N$ necessary to reach the condition
$\ell=s_{\rm cte}-(\ell_1 + \ell_2+ \dots + \ell_n)$ for any $n$ mixed together.}
\label{fig:testsFirstpoint-crop}
\end{figure}

 Figure~\ref{fig:testsFirstpoint-crop} (left panels) shows 
that the pdf of 
$\ell$ 
(see, Fig.~\ref{fig:StartingPoint-crop} for the symbols) is equal to the pdf of $\ell_n$ 
[Eq.~(\ref{eq:pdfTesselLength})].
Two representative cases for a small and a large $s_{\rm cte}$ are reported.
The data follow a straight lines due to the expected exponential behavior.
For a given   $s_{\rm cte}$, we also observe two different lines, 
depending  whether we consider tessels of type ``a'' or type ``b''.
  Hence, MC data for $\ell_n$ have the same pdf as $\ell$ (all possible $n$ are
mixed on the same line because the behavior is the same).
This means that, no matter where a photon starts inside a tessel,
Eq.~(\ref{eq:psFINALmix}) always remains valid.
This behavior remains the same for any choice of the parameters 
(simulations not  reported  here for obvious reasons).

The panels in the right column of Fig.~\ref{fig:testsFirstpoint-crop}
represent the histogram of  the number of tessels $N$ necessary to reach the condition
$\ell=s_{\rm cte}-(\ell_1 + \ell_2+ \dots + \ell_n)$ for any $n$.
We can see that, as expected,  if $s_{\rm cte}$ is small then the probability to have small $N$
is also low, and vice versa.

\section*{Conclusions}

In the present contribution we have derived the SSF [Eq.~(\ref{eq:psFINALmix})] for an 
isotropic Poisson tesselation with a binary mixture model.
The SSF allow us to perform MC simulations for any medium geometry, 
as for the classical Beer--Lambert--Bouguer case,  but where the tesselation 
is built ``on the flight''.
In other words, once the SSF is known, the binary medium can be treated as a homogeneous medium, and only one MC simulation is in principle necessary to obtain
the parameters of interest (instead of hundred of MC repetitions for each random tessel configuration).

Random $s$ based on the SSF Eq.~(\ref{eq:psFINALmix}) can be generated, e.g., as usual,
by deriving a look-up table relating $\xi$ and $s$ from
\begin{equation}
\xi=\int _0^s p_{s_{\rm mix}}(s';\mu_{t_a},\mu_{t_b},\sigma_a,\sigma_b) ds'.
\end{equation}

 The approach presented here for the derivation of the SSF may in principle be applied to other stochastic models.
 
We hope that the present contribution will give to the scientific community a further tool allowing to study photons (particles)  propagation in random media. 

\section*{Appendix A: Calculation details of F(s;a,b,N), Eq.~(\ref{eq:F}) in the main text}
In this appendix, using Laplace transform techniques, we provide technical details for the calculation of the integral (Eq.~(\ref{eq:F}))
\begin{align}
F(s;a,b,N) = \int_0^{s}p_{Tes_N}(s';\mu_{t_a},\mu_{t_b},\sigma_a,\sigma_b,N)ds'  \, ,
\end{align}
where
\begin{eqnarray} 
p_{Tes_N}(s;\mu_{t_a},\mu_{t_b},\sigma_a,\sigma_b,N)  \underset{N>1}{=}  \left\lbrace
  \begin{array}{lll}
   \frac{\sqrt{\pi }   (\mu_{t_b}+\sigma_b)^{\frac{N}{2}}(\mu_{t_a}+\sigma_a)^{\frac{N}{2}}   }{\mu^{\frac{N-1}{2}} 
  \Gamma \left(\frac{N}{2}\right)}
s^{\frac{N-1}{2}} I_{\frac{N-1}{2}}\left(\frac{s \mu }{2}\right)  e^{-\frac{(\mu_{t_a}+\mu_{t_b}+\sigma_a+\sigma_b)}{2}s}
    &N~\mathrm{even} 
    \\ 
\frac{\sqrt{\pi }   (\mu_{t_b}+\sigma_b)^{\frac{N-1}{2}}(\mu_{t_a}+\sigma_a)^{\frac{N+1}{2}}   }{2\mu^{\frac{N-2}{2}} 
  \Gamma \left(\frac{N+1}{2}\right)}
s^{\frac{N}{2}} 
\left[I_{\frac{N-2}{2}}\left(\frac{s \mu }{2}\right)+I_{\frac{N}{2}}\left(\frac{s \mu }{2}\right)\right]
  e^{-\frac{(\mu_{t_a}+\mu_{t_b}+\sigma_a+\sigma_b)}{2}s}
    &N~\mathrm{odd} \, .
  \end{array}
\right.
\end{eqnarray}
We were not able to perform the integral directly, instead we use Laplace transform techniques. We denote by $\mathcal{L}\{f(s)\} = \int_0^{+\infty} dt e^{-st } f(s)$ the Laplace transform of the function $f(s)$ and by $\mathcal{L}^{-1}\{f(t)\} = \int_{c-i\infty}^{c+i\infty} \frac{dt}{2 i \pi}  e^{s t} f(t)  $ its inverse Laplace transform. For sake of simplicity, we only treat the case where $N=2 p$ is even (the odd case can be handled in the same way). First, we get

\begin{align}
 \mathcal{L}\{F(s;a,b,2 p)\} &= \frac{\sqrt{\pi }   (\mu_{t_b}+\sigma_b)^{p}(\mu_{t_a}+\sigma_a)^{p}   }{\mu^{\frac{2p-1}{2}} 
  \Gamma \left(p \right)} \mathcal{L}\{\int_0^{s} 
s'^{\frac{2p-1}{2}} I_{\frac{2p-1}{2}}\left(\frac{s' \mu }{2}\right)  e^{-\frac{(\mu_{t_a}+\mu_{t_b}+\sigma_a+\sigma_b)}{2}s'} ds' \} \nonumber \\
						   &=  \frac{(\mu_a+\sigma_a)^p (\mu_b+\sigma_b)^p}{t (t + \mu_a+\sigma_a)^p (t + \mu_b+\sigma_b)^p}  \, ,
\end{align}
thus 
\begin{eqnarray}
 F(s;a,b,N) = (\mu_a+\sigma_a)^p (\mu_b+\sigma_b)^p  \mathcal{L}^{-1}\left \{\frac{1}{t (t + \mu_a+\sigma_a)^p (t + \mu_b+\sigma_b)^p} \right \}  \, .
\end{eqnarray}
We seek the inverse Laplace transform by decomposing the fraction $1/(t (t + \mu_a+\sigma_a)^p (t + \mu_b+\sigma_b)^p)$ into simple elements.
\begin{align}{}
  \frac{1}{t (t + \mu_a+\sigma_a)^p (t + \mu_b+\sigma_b)^p} & = \frac{1}{t \, (\mu_a+\sigma_a)^p  (\mu_b+\sigma_b)^p}  
  \nonumber \\
  &+
    \sum _{k=1}^{p} \frac{ \sum _{i=0}^{p-k} (-1)^{k+i+1} (\mu_a+\sigma_a)^i (\mu_b+\sigma_b)^{p-k-i} \binom{2 p-k}{i}}{(\mu_a+\sigma_a)^{p-k+1}  (\mu_a+\sigma_a-\mu_b-\sigma_b)^{2 p-k}} \frac{1}{(\mu_a+\sigma_a+t)^k} \nonumber \\
  &  + \sum _{k=1}^{p} \frac{ \sum _{i=0}^{p-k} (-1)^{k+i+1} (\mu_b+\sigma_b)^i (\mu_a+\sigma_a)^{p-k-i} \binom{2 p-k}{i}}{(\mu_b+\sigma_b)^{p-k+1}  (\mu_b+\sigma_b-\mu_a-\sigma_a)^{2 p-k}} \frac{1}{(\mu_b+\sigma_b+t)^k} .  \nonumber \\ 
\end{align}
One of the sums can be performed by resorting to the hypergeometric function $_2F_1(.,.;.;.)$:
\begin{align}
  \frac{(\mu_a+\sigma_a)^p (\mu_b+\sigma_b)^p}{t (t + \mu_a+\sigma_a)^p (t + \mu_b+\sigma_b)^p} = \frac{1}{t} 		& + \sum _{k=1}^{p} \frac{1}{(\mu_a+\sigma_a+t)^{k}} \left[ -(\mu_a+\sigma_a)^{k-1} +(-1)^{p+1} (\mu_a+\sigma_a-\mu_b-\sigma_b)^{k-2p} \right. \nonumber \\
  	& \left. \times (\mu_a+\sigma_a)^p b^{p-1} \binom{2 p-k}{p-k+1} \, _2F_1\left(1,1-p;p-k+2;\frac{\mu_a+\sigma_a}{\mu_b+\sigma_b}\right)\right] \nonumber \\
	& +\sum _{k=1}^{p} \frac{1}{(\mu_b+\sigma_b+t)^{k}} \left[ -(\mu_b+\sigma_b)^{k-1} +(-1)^{p+1} (\mu_b+\sigma_b-\mu_a-\sigma_a)^{k-2p} \right. \nonumber \\
  	& \left. \times (\mu_a+\sigma_a)^{p-1} (\mu_b+\sigma_b)^{p}\binom{2 p-k}{p-k+1} \, _2F_1\left(1,1-p;p-k+2;\frac{\mu_b+\sigma_b}{\mu_a+\sigma_a}\right)\right. \, . 
\end{align}
Since $\mathcal{L}^{-1}\left \{\frac{1}{t} \right \} = 1 $ and $\mathcal{L}^{-1}\left \{\frac{1}{(t + x)^k}  \right \} = \frac{e^{-x s} s^{k-1}}{\Gamma (k)}$ finally we obtain
\begin{align}
  F(s;a,b,2p) = 1 & + e^{-(\mu_a+\sigma_a) s} \sum _{k=1}^{p} \frac{s^{k-1}}{\Gamma (k)} \left[ -(\mu_a+\sigma_a)^{k-1} +(-1)^{p+1} (\mu_a+\sigma_a-\mu_b-\sigma_b)^{k-2p}\right. \nonumber \\
  	& \left. \times (\mu_a+\sigma_a)^p (\mu_b+\sigma_b)^{p-1} \binom{2 p-k}{p-k+1} \, _2F_1\left(1,1-p;p-k+2;\frac{\mu_a+\sigma_a}{\mu_b+\sigma_b}\right)\right] \nonumber \\
& +e^{-(\mu_b+\sigma_b) s} \sum _{k=1}^{p}  \frac{s^{k-1}}{\Gamma (k)} \left[-(\mu_b+\sigma_b)^{k-1} +(-1)^{p+1} (\mu_b+\sigma_b-\mu_a-\sigma_a)^{k-2p}\right. \nonumber \\
  	& \left. \times (\mu_a+\sigma_a)^{p-1} (\mu_b+\sigma_b)^{p}\binom{2 p-k}{p-k+1} \, _2F_1\left(1,1-p;p-k+2;\frac{\mu_b+\sigma_b}{\mu_a+\sigma_a}\right)\right] \, ,
\end{align}
which is the announced result Eq.(\ref{eq:Feven}) with $p = N/2$ and $\mu=\mu_{t_b}-\mu_{t_a} +\sigma_b-\sigma_a$.


%

%

\section*{Author contributions statement}

TB and AM contributed equally to this work. All authors reviewed the manuscript.

\section*{Additional information}

\textbf{Competing financial interests:} The authors declare no competing financial interests. 

\end{document}